\documentstyle[prd,eqsecnum,aps]{revtex}

\begin{document}

\draft
\title{Spinodal effect in the natural inflation model}
\author{Shinji Tsujikawa \thanks{electronic
address:shinji@gravity.phys.waseda.ac.jp}}
\address{ Department of Physics, Waseda University,
3-4-1 Ohkubo, Shinjuku-ku, Tokyo 169-8555, Japan\\[.3em]}
\author{Takashi Torii \thanks{electronic
address:torii@th.phys.titech.ac.jp}}
\address{Department of Physics, Tokyo
Institute of Technology, 2-12-1 Oh-Okayama Meguro, 
Tokyo 152-8551, Japan\\~}
\date{\today}
\maketitle
\begin{abstract}
Recently, Cormier and Holman pointed out that 
fluctuations of an inflaton field $\phi$ are significantly
enhanced in the model of {\it spinodal inflation} with a
potential $V(\phi)$ for which the second derivative
$V^{(2)}(\phi)$ changes sign.
As an application of this model, we investigate particle
production in the natural inflation model with a potential 
$V(\phi)=m^4 \left[1+\cos (\phi/f)\right]$
by making use of the Hartree approximation.
For typical mass scales $f \sim m_{\rm pl} 
\sim 10^{19}$GeV, and $m \sim
m_{\rm GUT} \sim 10^{16}$GeV, we find that 
growth of fluctuations relevantly occurs for the initial
value of inflaton $\phi(0)~\mbox{\raisebox{-1.ex}{$\stackrel
     {\textstyle<}{\textstyle \sim}$}}~0.1 m_{\rm pl}$. 
Especially for $\phi(0)~\mbox{\raisebox{-1.ex}{$\stackrel
     {\textstyle<}{\textstyle \sim}$}}~10^{-6}m_{\rm pl}$,
maximum fluctuations are so large that secondary 
inflation takes place by produced fluctuations.
In this case, the achieved number of $e$-folding becomes
much larger than in the case where an effect of spinodal
instability is neglected.
\end{abstract}


\baselineskip = 24pt

%
\section{Introduction}                            %
The concept of inflationary cosmology is very attractive 
to describe the early stage of the universe\cite{GS}.
It not only solves horizon, flatness, and monopole problems
the standard big bang cosmology has to face, but also
provides large-scale density perturbations required for
the structure formation\cite{KT}.
Inflation takes place while a scalar field $\phi$ called
{\it inflaton} slowly rolls down toward the minimum 
of its potential $V(\phi)$. 
The inflationary period ends when the inflaton field begins 
to oscillate around the minimum of its potential.
The elementary particles can be efficiently
produced by a nonperturbative process called 
{\it preheating}\cite{TB,KLS} in the oscillating 
stage of inflaton. Then these particles decay to other
lighter particles and thermalize the universe.

So far, there are various models of inflation.
These different kinds of models can be classified in the
following way~\cite{Kolb}.
The first class is the ``large field" model, in which 
the initial value of inflaton is large and rolls down toward
the potential minimum.
Chaotic inflation\cite{Chaotic} is one of 
the representative models of this class.
The second class is the ``small field" model, in which 
inflaton is small initially and slowly evolves toward
the potential minimum at larger values of $\phi$.
New inflation\cite{New} and natural inflation\cite{NI} 
are good examples of this model.
In the first model, the second derivative of the potential
$V^{(2)}(\phi)$ usually takes positive values, but in the second model,
$V^{(2)}(\phi)$ can change sign during inflation.
The third one is the hybrid inflation model\cite{hybrid}, 
where the inflaton field has a large amount of potential 
energy at the minimum of its potential, while the vacuum
energy is almost zero at the end of inflation in the first and
second inflation models.

Recently, Cormier and Holman\cite{SP} have pointed out 
that fluctuations of inflaton can grow nonperturbatively
during an inflationary stage in the second model of inflation
when $V^{(2)}(\phi)$ is negative.
This idea is remarkable in the sense that 
particles are effectively produced by negative instability
even in the slow rolling stage of inflaton.
They call this kind of model {\it spinodal inflation},
and investigated the nonperturbative evolution of
the inflaton field making use of the Hartree approximation
in the toy model with a potential
 $V(\phi)=\frac{3m^4}{2\lambda}
-\frac12 m^2\phi^2+\frac{\lambda}{4!}\phi^4$.
It was found that  low momentum 
modes of fluctuations are mainly enhanced, and
the evolution of the system can be 
described effectively by two classical scalar fields.
Especially when $\phi$ is close to zero initially,
the amount of the maximum fluctuation becomes so large that 
secondary inflation occurs by the effect of 
produced particles.
It was suggested that ordinary prediction of scale
invariance of density perturbations generated during inflation
would be modified by taking into account 
this {\it spinodal effect}.

As for one example of spinodal inflation, we draw
attention to the natural inflation model which was 
originally proposed by Freese et al.\cite{NI}.
This model is characterized by  pseudo 
Nambu-Goldstone bosons (PNGB) which appear
when an approximate global symmetry is 
spontaneously broken.
The PNGB potential is expressed as 
$V(\phi)=m^4 \left[1+\cos (\phi/f)\right]$,
where $f$ and $m$ are two mass scales
which characterize the shape of the potential.
Considering PNGB as the candidate of inflaton,
$f$ and $m$ are constrained by the requirement of
sufficient inflation and primordial density perturbations
observed by the Cosmic Background Explorer (COBE) satellite. 
In the case where the effect of spinodal instability
is neglected, these mass scales are found to be
$f \sim m_{\rm pl} \sim 10^{19}$GeV, and $m \sim
m_{\rm GUT} \sim 10^{16}$GeV respectively.
While other inflation models require extremely weak
coupling $\lambda$ in order to satisfy the constraint of 
density perturbations (for example, in the chaotic inflation
model with self coupling potential,
$\lambda~\mbox{\raisebox{-1.ex}{$\stackrel
     {\textstyle<}{\textstyle \sim}$}}~10^{-13}$), the PNGB inflation model 
is preferable in the sense that two mass scales arise naturally in 
particle physics models.
Furthermore, this model has an advantage 
in the analysis of the spinodal effect.
When fluctuations of inflaton grow 
significantly, higher order terms of fluctuations
play a relevant role for the evolution of the system.
As compared with other more complicated spinodal models 
such as new inflation, we can handle these higher order terms
in an analytic way in the natural inflation model.
What we are concerned with is how the efficient particle 
production during the natural inflation would modify
the dynamics of the system.
As we will show later, secondary inflation due to fluctuations
pointed out in Ref.~\cite{SP} appears and the evolution is 
drastically changed if the initial value
of inflaton is close to zero.

This paper is organized as follows.
In the next section, basic equations based on the 
Hartree approximation are introduced
in the natural inflation model.
In Sec.~III, we study how the fluctuation of inflaton is 
generated during the inflationary stage.
We will show that this effect can significantly alter
the dynamics of inflation.
We present our discussions and conclusions in the final section.

\section{Basic equations}   

The model we consider is 
\begin{eqnarray}
{\cal L} = \sqrt{-g} \left[ \frac{1}{2\kappa^2}R
   -\frac{1}{2}(\nabla \phi)^2
   -V(\phi) \right],
\label{B1}
\end{eqnarray}
where $\kappa^{2}/8\pi \equiv G =m_{\rm pl}^{-2} $ is Newton's
gravitational constant, $R$ is a scalar curvature, and 
$\phi$ is a minimally coupled inflaton field.
In this paper, we adopt a potential which is the so called 
natural inflation type
\begin{eqnarray}
V(\phi)=m^4 \left[1+\cos\left(\frac{\phi}{f}\right) \right],
\label{B2}
\end{eqnarray}
where two mass scales $m$ and $f$ characterize
the height and width of the potential, respectively. 
The typical mass scales are of the order
$f \sim m_{\rm pl} \sim 10^{19}$ GeV and 
$m \sim m_{\rm GUT}\sim 10^{16}$ GeV
for the success of the scenario of the natural
inflation\cite{NI}.
The inflaton field is initially located in the region of
$0 <\phi (0)< \pi f$, and 
inflation takes place while 
inflaton evolves slowly toward the minimum of
its potential at $\phi=\pi f$. 
In order to obtain the 
sufficient inflation by which the number of
$e$-folding exceeds $N \sim 60$, the initial value 
of inflaton is required to be close to $\phi=0$.
For example, in the case of 
$f=10^{19}$ GeV,
we need $\phi(0)~\mbox{\raisebox{-1.ex}{$\stackrel
     {\textstyle<}{\textstyle \sim}$}}~0.5 m_{\rm pl}$.
The value of $\phi$
when inflation ends ($=\phi_F$) depends 
on the scale $f$, but $\phi_F$ is close to 
the value $\pi f$ for the typical values of 
$f \sim m_{\rm pl}$\cite{NI}.
When inflaton begins to oscillate around $\phi
=\pi f$, the system enters a reheating stage.

For the potential $(\ref{B2})$, we find that 
$V^{(2)}(\phi)$ is negative when  inflaton evolves in the region
of $0 <\phi < \pi f/2$.
This leads to the enhancement of fluctuations
by spinodal instability in the realistic initial value of 
$\phi$. After the inflaton field passes through
$\phi=\pi f/2$ where $V^{(2)}(\phi)$ changes sign,
fluctuations of inflaton no longer grow.
Hence an important point for the development 
of fluctuations is the initial value of inflaton.
The mass scales $f$ and $m$
also affect the evolution of the system.

Let us obtain basic equations in the natural inflation model.
We adopt the flat Friedmann-Robertson-Walker metric
\begin{eqnarray}
ds^2 = -dt^2 + a^2(t) d {\bf x}^2,
\label{B3}
\end{eqnarray}
where $a(t)$ is the scale factor, and $t$ is the 
cosmic time coordinate. 

We decompose the quantum scalar field $\phi(t,{\bf x})$ into
its expectation value $\phi_0(t)$ and the quantum fluctuation
$\delta \phi(t,{\bf x})$ as
\begin{eqnarray}
\phi(t,{\bf x})=\phi_0(t) +\delta \phi(t,{\bf x}),
\label{B4}
\end{eqnarray}
with 
\begin{eqnarray}
\phi_0(t) = \langle \phi(t,{\bf x}) \rangle
=\frac{{\rm Tr} \phi \rho(t)}{{\rm Tr} \rho(t)},
\label{B200}
\end{eqnarray}
where $\rho(t)$ is the density matrix which satisfies 
the Liouville equation:
\begin{eqnarray}
i \frac{d\rho(t)}{dt}=[{\cal H},\rho(t)].
\label{B201}
\end{eqnarray}
${\cal H}$ is the time dependent Hamiltonian.
The initial condition for the density matrix should be
chosen to describe a local thermal equilibrium state.
Given the initial condition, the time
evolution of $\rho(t)$ is known from Eq.~$(\ref{B201})$.
The expectation value of the energy-momentum tensor is
evaluated from the time varying density matrix.
Our approach is to solve the semiclassical Einstein equations
where the source of the gravitational field is 
the expectation value of the energy-momentum tensor .

The system we consider is out-of-equilibrium 
and nonperturbative,
and there are some approximations which are suitable 
to describe such a state.
In this paper, we adopt the Hartree mean field approximation
of the nonequilibrium quantum field theory.
This is basically a Gaussian variational approximation 
to the time dependent density matrix.
Related with this issue, several authors\cite{HABIB,Boy1,Boy2} considered 
the large $N$ approximation which can deal with contributions beyond 
leading order. Although it is of interest to examine the differences 
between two approximations, the analysis based on the
large $N$ approximation is left for the future work\cite{comment}.

Performing the Hartree factorization
\begin{eqnarray}
\delta \phi^{2n} &\to& \frac{(2n)!}{2^n (n-1)!} 
\langle \delta \phi^2 \rangle^{n-1} \delta\phi^2-
\frac{(2n)! (n-1)}{2^n n!} \langle \delta \phi^2 \rangle^n, 
\nonumber \\
\delta \phi^{2n+1} &\to& \frac{(2n+1)!}{2^n n!}
\langle \delta \phi^2 \rangle^n \delta\phi,
\label{B300}
\end{eqnarray}
and making use of the tadpole condition
\begin{eqnarray}
\langle \delta \phi(t,{\bf x}) \rangle=0,
\label{B5}
\end{eqnarray}
the expectation value of the potential $V(\phi_0+\delta\phi)$
can be written as
\begin{eqnarray}
\langle V(\phi_0+\delta\phi) \rangle=
\sum_{n=0}^{\infty} \frac{1}{2^n n!}
\langle \delta\phi^2 \rangle^n V^{(2n)}(\phi_0),
\label{B301}
\end{eqnarray}
where $V^{(n)}(\phi) \equiv  \delta^n V(\phi)/\delta \phi^n$.
Then the equation of the $\phi_0$ field yields
\begin{eqnarray}
\ddot{\phi}_0 +3H \dot{\phi}_0 +
\sum_{n=0}^{\infty} \frac{1}{2^n n!}
\langle \delta\phi^2 \rangle^n V^{(2n+1)} (\phi_0) 
=0,
\label{B7}
\end{eqnarray}
where a dot denotes a derivative with respect to the 
cosmic time coordinate, $H \equiv \dot{a}/a$ is 
the Hubble parameter.
Expanding the $\delta \phi$ field by the Fourier 
modes as
\begin{eqnarray}
\delta \phi=\frac{1}{(2\pi)^{3/2}} \int 
\left(a_k 
\delta \phi_k(t)
 e^{-i {\bf k} \cdot {\bf x}}+a_k^{\dagger} 
\delta \phi_k^{*}(t)
 e^{i {\bf k} \cdot {\bf x}} 
\right) d^3{\bf k},
\label{B6}
\end{eqnarray}
we obtain the following equation for the fluctuation: 
\begin{eqnarray}
\delta \ddot{\phi}_k +3H \delta \dot{\phi}_k+
\left[ \frac{k^2}{a^2} +\sum_{n=0}^{\infty} \frac{1}{2^n n!}
\langle \delta\phi^2 \rangle^n V^{(2n+2)} (\phi_0) 
\right] \delta \phi_k=0,
\label{B8}
\end{eqnarray}
where the expectation value $\langle \delta \phi^2 \rangle$
is represented by
\begin{eqnarray}
\langle \delta \phi^2 \rangle = \frac1{2\pi^2} \int k^2
|\delta \phi_k|^2 dk.
\label{B9}
\end{eqnarray}
The evolution of the scale factor is written as
\begin{eqnarray}
\left(\frac{\dot{a}}{a}\right)^2 =
   \frac{\kappa^2}{3} 
     \biggl[ \frac12 \dot{\phi_0}^2
   +\frac12 \langle \delta \dot{\phi}^2 \rangle
   +\frac{1}{2a^2} \langle (\nabla \delta\phi )^2
    \rangle 
   +\sum_{n=0}^{\infty} \frac{1}{2^n n!}
  \langle \delta\phi^2 \rangle^n V^{(2n)}(\phi_0)
\biggr],
\label{B11}
\end{eqnarray}
where $\langle \delta \dot{\phi}^2 \rangle$ and
$\langle (\nabla \delta\phi )^2 \rangle$ 
are expressed by
\begin{eqnarray}
\langle \delta \dot{\phi}^2 \rangle=
\frac1{2\pi^2} \int k^2
|\delta \dot{\phi}_k|^2 dk,
\label{B12}
\end{eqnarray}
\begin{eqnarray}
\langle (\nabla \delta\phi )^2 \rangle =
\frac{1}{2\pi^2} \int k^4
|\delta \phi_k|^2 dk.
\label{B13}
\end{eqnarray}
The quantities of Eqs.~$(\ref{B9})$, $(\ref{B12})$, and 
$(\ref{B13})$ need to be regulated in order to remove
the divergences of integrals.
Several authors considered renormalizations 
by the method of adiabatic 
regularization\cite{Boy1,Boy2,PRA,RH} and 
dimensional regulalization\cite{Baa}.
The former scheme is based on introducing a large momentum
cutoff and subtracting the leading adiabatic orders of the 
fluctuation terms. 
The latter is the covariant regularization in which 
the counter terms do not depend on the initial state.
However, this dimensional regulalization has a shortcoming
that the energy-momentum tensor has an initial singularity.
In this paper, we make use of the scheme of adiabatic regularization
as in Ref.~\cite{Boy1,Boy2}, which is suitable for numerical 
computations.

In the natural inflation potential  $(\ref{B2})$, Eqs. $(\ref{B7})$, $(\ref{B8})$,
and $(\ref{B11})$ can be rewritten as
\begin{eqnarray}
\ddot{\phi}_0 +3H \dot{\phi}_0 -\frac{m^4}{f}
F(\langle\delta \phi^2\rangle)  \sin 
\left( \frac{\phi_0}{f} \right)
=0,
\label{B14}
\end{eqnarray}
\begin{eqnarray}
\delta \ddot{\phi}_k +3H \delta \dot{\phi}_k+
\left[ \frac{k^2}{a^2} -\frac{m^4}{f^2}
F(\langle\delta \phi^2\rangle) \cos  
\left(\frac{\phi_0}{f} \right)
\right] \delta \phi_k=0,
\label{B15}
\end{eqnarray}
\begin{eqnarray}
\left(\frac{\dot{a}}{a}\right)^2 = 
   \frac{\kappa^2}{3} 
     \left\{ \frac12 \dot{\phi_0}^2
   +\frac12 \langle \delta \dot{\phi}^2 \rangle
   +\frac{1}{2a^2} \langle (\nabla \delta\phi )^2
   \rangle +
  m^4 \left[ 1+F(\langle\delta \phi^2\rangle) 
  \cos \left( \frac{\phi_0}{f}\right) 
  \right] \right\},
\label{B16}
\end{eqnarray}
where
\begin{eqnarray}
F(\langle\delta \phi^2\rangle) \equiv \exp 
\left(-\frac{\langle\delta \phi^2\rangle}{2f^2} \right).
\label{B17}
\end{eqnarray}
Note that $F(\langle\delta \phi^2\rangle)=1$
in the case of $\langle\delta \phi^2\rangle=0$.
When $\phi_0$ is located in the region $0<\phi_0<\pi f/2$
initially,  the term in the square bracket in Eq.~$(\ref{B15})$
is negative for small $k$ at the first stage of inflation.
This leads to the enhancement of fluctuations of low
momentum modes.
Since $\phi_0$ increases for $0<\phi_0<\pi f$
as is found in Eq.~$(\ref{B14})$,
the term $\cos (\phi_0/f)$ in Eq.~$(\ref{B15})$ 
gradually decreases 
and the growth of fluctuations terminates 
after $\phi_0$ becomes larger than $\pi f/2$.

As is found by Eq.~$(\ref{B16})$,
the evolution of inflaton can be described effectively
by two homogeneous fields 
with the potential
\begin{eqnarray}
 V(\phi_0, \sigma) \equiv
  m^4 \left[ 1+ \exp \left(-\frac{\sigma^2}{2f^2}
  \right) \cos  \left(\frac{\phi_0}{f}\right) \right], 
\label{B18}
\end{eqnarray}
where $\sigma \equiv \sqrt{\langle\delta \phi^2\rangle}$.
This potential is depicted in Fig.~1.
When fluctuations do not grow 
relevantly as $\sigma \ll f$, the $\phi_0$ field slowly rolls
down toward the potential minimum $\phi_0=\pi f$
almost along the $\phi_0$ direction
in the usual manner.
The inflationary period ends when the $\phi_0$ field begins
to oscillate around the minimum of its potential.
At this stage, the potential energy becomes small and 
the expansion rate slows down. 
However, in the case where $\sigma$ grows to the 
order of $f$, the evolution of the system is drastically modified.
The inflaton field moves toward the $\sigma$ direction
rather than the $\phi_0$ direction in Fig.~1.
It reaches the flat region $\sigma~\mbox{\raisebox{-1.ex}{$\stackrel
     {\textstyle>}{\textstyle\sim}$}}~2m_{\rm pl}$, and 
secondary inflation occurs there.
We investigate for what initial values of $\phi_0$ this behavior
appears in the next section.

Before analyzing the evolution of the system, we mention
the initial conditions for the fluctuation.
We should choose a conformal adiabatic vacuum state where the density matrix 
represents a local thermal equilibrium,
which means that the density matrix commutes with 
the initial conformal Hamiltonian.
This corresponds to choose the mode functions 
$\delta \phi_k$ as
\begin{eqnarray}
\delta \phi_k(0) = \frac{1}{\sqrt{2\omega_k (0)}}, ~~~
\delta \dot{\phi}_k(0) = 
\left[-i \omega_k(0) -H(0) \right] \delta \phi_k(0),
\label{B21}
\end{eqnarray}
with 
\begin{eqnarray}
\omega_k^2(0) = k^2+{\cal M}^2 (0),~~~~
{\cal M}^2 (0) = -\frac{m^4}{f^2}
F(\langle\delta \phi^2\rangle)
 \cos \left( \frac{\phi_0(0)}{f} \right)
-\frac{R(0)}{6}, 
\label{B23}
\end{eqnarray}
where $R(0)$ is the initial scalar curvature, and 
we set  $a(0)=1$.

In the present model, since $\omega_k^2$ becomes negative 
for small $k$, we need to modify the initial frequencies of 
low momentum modes.
Following  the approach performed in Ref.~\cite{Boy1},
we adopt the initial frequency as
\begin{eqnarray}
\omega_k^2(0)=k^2+{\cal M}^2 (0)
~{\rm tan h}\left(\frac{k^2+{\cal M}^2 (0)}
{|{\cal M}^2 (0)|} \right),
\label{B22}
\end{eqnarray}
This initial frequency coincides with the 
conformal vacuum frequency $(\ref{B23})$ for large $k$,
and becomes positive for small $k$.
This choice of the initial frequency  smoothly interpolates
between large and small momentum modes. 
An alternative way is to choose the initial condition as\cite{Boy2}
\begin{eqnarray}
\omega_k^2(0) = \cases{
 k^2+|{\cal M}^2 (0)| & {\rm with} $k^2<|{\cal M}^2 (0)|$, \cr
 k^2+{\cal M}^2 (0) & {\rm with} $k^2 \ge |{\cal M}^2 (0)|$. \cr
 }
\label{B60}
\end{eqnarray}
Although there are some subtleties about the choice of 
initial frequencies,
we can numerically check that these different choices 
have little effect on the evolution of the system.
The qualitative properties of the system are the same
in either case of Eq.~$(\ref{B22})$ or Eq.~$(\ref{B60})$.

We investigate the nonperturbative evolution of
$\langle\delta \phi^2\rangle$ with the initial condition of 
Eqs.~$(\ref{B21})$ and $(\ref{B22})$
as the semiclassical problem.

\section{Particle production in the natural inflation model}   

In this section, we study out-of-equilibrium dynamics due to 
the enhancement of fluctuations in the natural inflation model.
Let us first consider the typical case of 
$f=10^{19}$ GeV$ \sim m_{\rm pl}$
and $m=10^{16}$ GeV$ \sim 10^{-3}m_{\rm pl}$.
In order to solve the standard cosmological puzzles
of the standard big bang cosmology, the 
number of $e$-folding
\begin{eqnarray}
N \equiv {\rm ln} \left( \frac{a(t_f)}{a(0)} \right),
\label{C1}
\end{eqnarray}
where $t_f$ denotes the time when the slow roll period ends,
is required to be  $N~\mbox{\raisebox{-1.ex}{$\stackrel
     {\textstyle>}{\textstyle\sim}$}}~60$.
We need initial values of inflaton as  
$\phi(0)~\mbox{\raisebox{-1.ex}{$\stackrel
     {\textstyle<}{\textstyle \sim}$}}~0.5 m_{\rm pl}$ to obtain 
 $N~\mbox{\raisebox{-1.ex}{$\stackrel
     {\textstyle>}{\textstyle\sim}$}}~60$
in the case where the spinodal effect is not included\cite{NI}. 
For examples, when $\phi(0)=0.5 m_{\rm pl}$, $N=71$; 
and when $\phi(0)=m_{\rm pl}$, $N=39$.
One may consider that the enhancement of fluctuations would
lead to the larger amount of $e$-folding and relax the constraint
of $\phi(0)$ to yield $N~\mbox{\raisebox{-1.ex}{$\stackrel
 {\textstyle>}{\textstyle\sim}$}}~60$.
However, this is not the case. 
 Fluctuations do not grow relevantly
 for $\phi(0)~\mbox{\raisebox{-1.ex}
{$\stackrel{\textstyle>}{\textstyle\sim}$}}~
0.1 m_{\rm pl}$.
We depict the evolution of the $\phi_0$ field and the 
fluctuation $\sigma$ for the case of 
$\phi(0)=0.1 m_{\rm pl}$ in Fig.~2.
We find that the maximum fluctuation  at
$\phi_0=\pi f/2$ is 
$\sigma_{\rm max} \approx 10^{-5}m_{\rm pl}$.
Since $\sigma_{\rm max}^2/(2f^2) \ll 1$ and 
$F(\langle\delta \phi^2\rangle)$ is close to unity, 
the evolution of the $\phi_0$ field and the scale 
factor are almost the same as in the case where 
the growth of the fluctuation is neglected.
The $\phi_0$ field evolves toward the potential
minimum at $\phi_0=\pi f$ without being 
affected by the back reaction effect of produced
particles. Inflationary period ends when
$mt \approx 4.5 \times 10^{4}$, at which 
the value of $\phi_0$ is $\phi_0 \approx 3.0f$
\cite{NI}.

In the case of $\phi(0)~\mbox{\raisebox{-1.ex}{$\stackrel
     {\textstyle<}{\textstyle \sim}$}}~0.1 m_{\rm pl}$,
the number of $e$-folding becomes larger 
with the decrease of $\phi(0)$ since 
the slow roll period is longer.
In addition to this, fluctuations are enhanced
more efficiently.
In TABLE I, we show numerical values of $N$ and 
$\sigma_{\rm max}$ in various cases of $\phi(0)$.
The number of $e$-folding $N'$ where the spinodal
effect is neglected is also presented.
We find that both $N$ and $\sigma_{\rm max}$ 
increase with the decrease of $\phi(0)$ 
for $10^{-6}m_{\rm pl}~\mbox{\raisebox{-1.ex}{$\stackrel
     {\textstyle<}{\textstyle \sim}$}}~\phi(0)~
     \mbox{\raisebox{-1.ex}{$\stackrel
     {\textstyle<}{\textstyle \sim}$}}~
10^{-1}m_{\rm pl}$.
The growth of fluctuations continues until the $\phi_0$
field reaches $\phi_0=\pi f/2$.
With the decrease of $\phi(0)$, since the period 
during which the inflaton field moves in the region
of $V^{(2)}(\phi)<0$ becomes longer, this results in the 
larger amount of the maximum fluctuation.
In the case of $10^{-6}m_{\rm pl}~
\mbox{\raisebox{-1.ex}{$\stackrel
     {\textstyle<}{\textstyle \sim}$}}~\phi(0)~
 \mbox{\raisebox{-1.ex}{$\stackrel
     {\textstyle<}{\textstyle \sim}$}}~
10^{-1}m_{\rm pl}$, 
numerical calculations show that $\sigma_{\rm max}$
can be approximately written as a function of $\phi(0)$:
\begin{eqnarray}
\sigma_{\rm max} \approx \left(\frac{\phi(0)}{10^{-6}m_{\rm pl}}
\right)^{-1} m_{\rm pl}.
\label{C2}
\end{eqnarray}
When $\phi(0)~\mbox{\raisebox{-1.ex}{$\stackrel
     {\textstyle>}{\textstyle\sim}$}}~10^{-5} m_{\rm pl}$,  
$\sigma_{\rm max}$ does not exceed $0.1 m_{\rm pl}$.
In this case, since $\langle\delta \phi^2\rangle/2f^2 \ll 1$ in 
Eq.~$(\ref{B17})$, the back reaction effect due to 
fluctuations can be neglected.
Although the inflaton field moves a bit toward
the $\sigma$ direction in Fig.~1, 
it slowly rolls down toward the minimum 
of its potential in the usual manner.
As a result, the number of $e$-folding does not 
change even taking into account the spinodal effect
(see TABLE I).

On the other hand, when $\phi(0)~\mbox{\raisebox{-1.ex}{$\stackrel
     {\textstyle<}{\textstyle \sim}$}}~10^{-6} m_{\rm pl}$, 
the fluctuation reaches $\sigma_{\rm max}
~\mbox{\raisebox{-1.ex}{$\stackrel
     {\textstyle>}{\textstyle\sim}$}}~m_{\rm pl}$. 
In this case, produced fluctuations play a relevant role 
for the evolution of the system.
For example, let us consider the case of  
$\phi(0)=5.0 \times 10^{-7} m_{\rm pl}$.
As is shown in Fig.~3, the fluctuation reaches the maximum 
value $\sigma_{\rm max} \approx 2.3 m_{\rm pl}$ 
at $mt=2.5 \times 10^5$, where
$\phi_0=\pi f/2$. After that, particle production 
terminates completely because $V^{(2)}(\phi)$ changes sign, 
and $\sigma$ decreases 
by the expansion of the universe.
We can see this behavior of inflaton in Fig.~1.
In this case, the inflaton field evolves toward the $\sigma$ direction
rather than the $\phi_0$ direction, and reaches the region around 
$\sigma \approx 2.3m_{\rm pl}$ and  
$\phi_0 \approx 1.5m_{\rm pl}$.
Since this region of the effective potential $(\ref{B18})$ 
is flatter than the region of 
$\sigma \approx 0$ and $\phi_0 \approx 1.5m_{\rm pl}$,
the inflaton field moves slowly
for some time. After that, it evolves along the valley around 
$0<\sigma/m_{\rm pl}<2.3$ and 
$\phi_0 \approx \pi f$, and finally arrives at the minimum of 
its potential. The amount of inflation is larger than 
in the case where the spinodal effect is ignored
as is found in TABLE I, because
produced fluctuations provide the additional energy density in 
Eq.~$(\ref{B16})$. 

This tendency becomes stronger with the decrease of $\phi(0)$.
In Fig.~4, we depict the evolution of $\phi_0$ and $\sigma$ 
fields in the case of $\phi_0(0)=3.0 \times 10^{-7} m_{\rm pl}$.
In this case, the fluctuation reaches the maximum
value $\sigma_{\rm max} \approx 3.8 m_{\rm pl}$.
One of the different points from 
the $\phi(0)=5.0 \times 10^{-7} m_{\rm pl}$ case is that 
the inflaton field stays in flat regions for a long time:
$1 \times 10^{6}~\mbox{\raisebox{-1.ex}{$\stackrel
     {\textstyle<}{\textstyle \sim}$}}~mt~
     \mbox{\raisebox{-1.ex}{$\stackrel
     {\textstyle<}{\textstyle \sim}$}}~7 \times 10^{6}$.
As is seen in Fig.~1, the effective potential $V(\phi_0,\sigma)$
in the region around 
$\sigma \approx 3.8m_{\rm pl}$ and 
$\phi_0 \approx 1.5m_{\rm pl}$ is very flat.
Since $F(\langle\delta \phi^2\rangle_{\rm max})$
is much smaller than unity, $V(\phi_0,\sigma)$ takes 
the almost constant value $m^4$.
The third terms in Eqs.~$(\ref{B14})$ and 
$(\ref{B15})$ become very small in this region
(since the main contributions to the fluctuation are due to
low momentum modes, $k^2/a^2 \to 0$),
and the inflaton field evolves very 
slowly in the flat region $V(\phi_0,\sigma) \approx m^4$.
This results in the secondary inflation supported by 
fluctuations. This behavior was originally pointed out by
Cormier and Holman in the model with a potential
$V(\phi)=\frac{3m^4}{2\lambda}
-\frac12 m^2\phi^2+\frac{\lambda}{4!}\phi^4$\cite{SP}.
We can expect that their results 
based on the Hartree approximation
generally hold in more complex spinodal type potentials 
if we choose initial values of inflaton close to zero.
When $\phi(0)=3.0 \times 10^{-7} m_{\rm pl}$,
the secondary inflation continues much longer than 
the first inflation
driven by the potential energy around $\phi=0$ at
$0~\mbox{\raisebox{-1.ex}{$\stackrel
     {\textstyle<}{\textstyle \sim}$}}~mt
     ~\mbox{\raisebox{-1.ex}{$\stackrel
     {\textstyle<}{\textstyle \sim}$}}
 ~1.3 \times 10^{5}$ (See Fig.~5).
This means that the number of $e$-folding is modified
to be much larger than in the case where particle production 
due to spinodal instability is neglected.
In fact, as is found in TABLE I, the number of $e$-folding
is very large as $N=22178$.
After the secondary inflation ends, the inflaton field rolls
down toward the potential minimum around $\phi_0=\pi f$
and $\sigma=0$, after which the universe enters the reheating
stage.

With the decrease of initial values of inflaton as
$\phi(0)~\mbox{\raisebox{-1.ex}{$\stackrel
     {\textstyle<}{\textstyle \sim}$}}~10^{-7}m_{\rm pl}$, 
since the fluctuation grows as 
$\sigma~\mbox{\raisebox{-1.ex}{$\stackrel
     {\textstyle>}{\textstyle\sim}$}}~4m_{\rm pl}$,
inflaton reaches the further flat region of the potential 
$V(\phi_0,\sigma)$. The duration of the secondary 
inflation becomes very long, and 
the amount of inflation is enormously large.
As long as $0<\phi_0<\pi f/2$, the inflaton field is to
roll down toward the $\sigma$ direction, and the secondary
inflation continues.
However, once $\phi_0$ exceeds the value of 
$\phi_0=\pi f/2$, $\sigma$ begins to decrease toward
$\sigma=0$.
The secondary inflation never ends in the extreme
case of $\phi_0=0$ as was pointed out in Ref.~\cite{SP}.

Next, we investigate the case where $m$ and $f$ are changed.
The mass $m$ is constrained by density perturbations observed
by COBE. The analytic estimation neglecting particle 
production due to spinodal instability\cite{NI} shows that 
$m$ is constrained by a function of $f$ as 
\begin{eqnarray}
m = \frac{1.7 \times 10^{16}}{b^{1/2}}
\left[ \frac{m_{\rm pl}}{f} \sin \left(\frac{\phi(t_f)}
{2f} \right) \right]^{1/2}
\exp\left(-\frac{15m_{\rm pl}^2}{8\pi f^2} \right)~
{\rm GeV},
\label{C3}
\end{eqnarray}
where $b$ is an overall bias factor which ranges 
$0.7<b<1.3$.
The term $\sin (\phi(t_f)/2f)$ 
is typically of the order of unity.
In the case where $f$ is of order $m_{\rm pl}$,
$m$ ranges in the region of $10^{15} {\rm GeV}
~\mbox{\raisebox{-1.ex}{$\stackrel
     {\textstyle<}{\textstyle \sim}$}}
     ~m~\mbox{\raisebox{-1.ex}{$\stackrel
     {\textstyle<}{\textstyle \sim}$}}~10^{16} {\rm GeV}$.
When $f$ is smaller than $m_{\rm pl}$ 
by one order of magnitude,
$m$ decreases significantly because the exponential term 
in Eq.~$(\ref{C3})$ plays a dominant role to determine
the mass scale.
If we include the effect of spinodal instability, since this would
produce the spatial inhomogeneity, the constraint $m$ will
be changed. In order to study this issue appropriately, we have 
to investigate the evolution of metric perturbations
during inflation. In the present model, however, since  
metric perturbations may be enhanced up to the 
nonlinear level by spinodal instability, the first order 
perturbation will not give the correct description of physics.
Although we do not consider these complex 
issues in this paper, it would be necessary to include 
metric perturbations for a complete study of nonperturbative dynamics.

Let us study the growth of fluctuations by changing the scale
$m$ for the fixed value of $f=10^{19}$ GeV$\sim m_{\rm pl}$.
Consider the case of $m=10^{15}$ GeV$\sim 10^{-4}m_{\rm pl}$.
When particle creation by spinodal instability is neglected, since 
the achieved number of $e$-folding is expressed by
\begin{eqnarray}
N=\frac{16\pi f^2}{m_{\rm pl}^2}
{\rm ln} \left[ \frac{\sin (\phi(t_f)/2f)}
{\sin (\phi(0)/2f)} \right],
\label{C4}
\end{eqnarray}
it does not depend on the scale of $m$.
Although it takes more time to terminate inflation 
for smaller values of $m$, we obtain 
$N~\mbox{\raisebox{-1.ex}{$\stackrel
     {\textstyle>}{\textstyle\sim}$}}~60$ for
$\phi(0)~\mbox{\raisebox{-1.ex}{$\stackrel
     {\textstyle<}{\textstyle \sim}$}}~0.5m_{\rm pl}$, 
which is the same as in the $m=10^{16}$ GeV case.
As for fluctuations of inflaton, since
$\langle \delta\phi^2\rangle$ is normalized by the square of
mass $m$, the achieved maximum value of $\sigma$ becomes
smaller as $m$ decreases for the same initial value of $\phi$.
Particle production relevantly occurs for the case of 
$\phi(0)~\mbox{\raisebox{-1.ex}{$\stackrel
     {\textstyle<}{\textstyle \sim}$}}~10^{-3} m_{\rm pl}$.
Numerical calculations show that the maximum fluctuation 
$\sigma_{\rm max}$ in the case of 
$10^{-8} m_{\rm pl}~\mbox{\raisebox{-1.ex}{$\stackrel
     {\textstyle<}{\textstyle \sim}$}}~\phi(0)
~\mbox{\raisebox{-1.ex}{$\stackrel
     {\textstyle<}{\textstyle \sim}$}}~10^{-3} m_{\rm pl}$ is smaller 
by two orders of magnitude than 
in the case of $m=10^{16}$ GeV (see TABLE II).
Namely, we find the relation 
\begin{eqnarray}
\sigma_{\rm max} \approx \left(\frac{\phi(0)}{10^{-6}m_{\rm pl}}
\right)^{-1} 10^{-2} m_{\rm pl}.
\label{C100}
\end{eqnarray}
When $\phi(0)~\mbox{\raisebox{-1.ex}{$\stackrel
     {\textstyle<}{\textstyle \sim}$}}~10^{-8}m_{\rm pl}$, 
$\sigma_{\rm max}$ exceeds the order of $f\sim m_{\rm pl}$,
and the secondary inflation occurs as in the case of
$m=10^{16}$ GeV. The number of 
$e$-folding becomes larger than in the case where
the spinodal effect is neglected,  
and it depends on the scale of $m$.
We found that the smaller values of $\phi(0)$ are required for the
development of fluctuations with the decrease of $m$.

Finally we comment on the case where the mass $f$
is changed. The number of $e$-folding is smaller 
with the decrease of $f$, because the potential $(\ref{B2})$
becomes steeper. Consider the case of $f=5.0 \times 10^{18}$ GeV
$\sim 0.5m_{\rm pl}$ and
$m=10^{15}$ GeV$\sim 10^{-4}m_{\rm pl}$. 
Even in the initial value of 
$\phi(0)=10^{-1}m_{\rm pl}$, the number of $e$-folding is 
only $N=33$.
In order to lead to the sufficient inflation as 
$N~\mbox{\raisebox{-1.ex}{$\stackrel
     {\textstyle>}{\textstyle\sim}$}}~60$,
we require initial values as 
$\phi(0)~\mbox{\raisebox{-1.ex}{$\stackrel
     {\textstyle<}{\textstyle \sim}$}}~10^{-2}m_{\rm pl}$. 
Since the inflaton field rolls down rapidly in the regions
of $V^{(2)}(\phi)<0$ compared with 
the case of $f=10^{19}$ GeV, the growth of fluctuations is 
slower. For example, the maximum fluctuations are 
$\sigma_{\rm max}=6.7 \times 10^{-4}m_{\rm pl}$ for 
$\phi(0)=10^{-5}m_{\rm pl}$; 
$\sigma_{\rm max}=6.7 \times 10^{-2} m_{\rm pl}$ for 
$\phi(0)=10^{-7}m_{\rm pl}$ (see TABLE III).
These values are smaller than in the case of 
$f=10^{19}$ GeV and $m=10^{15}$ GeV
for the same initial values of $\phi$.
The fluctuation grows up to the nonlinear level 
for $\phi(0)~\mbox{\raisebox{-1.ex}{$\stackrel
     {\textstyle<}{\textstyle \sim}$}}~5\times 10^{-9}m_{\rm pl}$, and 
the secondary inflation also occurs in this case.
For the values of $f$ which are not much smaller than the
Planck order, we can say that fluctuations are enhanced 
beyond the perturbative level and can support the total
amount of inflation.

\section{Concluding remarks and discussions}   

In this paper we have investigated the evolution
of an inflaton field $\phi$ in the presence of nonperturbative
behavior of fluctuations due to 
spinodal instability in the natural inflation model.
Since the second derivative $V^{(2)}(\phi)$ of the potential 
$V(\phi)=m^4 \left[1+\cos (\phi/f)\right]$ is negative
for the values of $0 <\phi<\pi f/2$, fluctuations of inflaton
can grow even during the inflationary phase.

The strength of the excitation of fluctuations $\sigma$ strongly 
depends on the initial value of inflaton $\phi(0)$.
For typical mass scales $f=m_{\rm pl}$ and $m=10^{-3}m_{\rm pl}$,
we have examined the dynamics of the system 
in various values of $\phi(0)$ by making use of 
the Hartree approximation.
For the values of $\phi(0)~\mbox{\raisebox{-1.ex}{$\stackrel
     {\textstyle<}{\textstyle \sim}$}}~0.5m_{\rm pl}$, we have
sufficient inflation as $N~\mbox{\raisebox{-1.ex}{$\stackrel
     {\textstyle>}{\textstyle\sim}$}}~60$ 
which is required to solve cosmological puzzles
of the big bang cosmology.
When $\phi(0)~\mbox{\raisebox{-1.ex}{$\stackrel
     {\textstyle<}{\textstyle \sim}$}}~0.1m_{\rm pl}$, 
fluctuations are relevantly enhanced with the decrease of $\phi(0)$ 
because duration of spinodal instability becomes longer.
Since long wavelength modes of fluctuations 
are mainly enhanced, the system can be described effectively 
by two homogeneous fields with potential $(\ref{B18})$.
The natural inflation model has the advantage that  higher 
order terms of fluctuations can be handled in an analytic way.
With the increase of $\sigma$,
the term $F(\langle\delta \phi^2\rangle)$ in Eq.~$(\ref{B17})$ 
decreases from unity.
This changes the evolution of the system as is found
in Eqs.~$(\ref{B14})$-$(\ref{B16})$.
Numerical calculations show that the maximum value
of fluctuations is $\sigma_{\rm max} \approx 
\left(\phi(0)/10^{-6}m_{\rm pl}
\right)^{-1} m_{\rm pl}$ for the case of 
$10^{-5}m_{\rm pl}~\mbox{\raisebox{-1.ex}{$\stackrel
     {\textstyle<}{\textstyle \sim}$}}~\phi(0)~
\mbox{\raisebox{-1.ex}{$\stackrel
     {\textstyle<}{\textstyle \sim}$}}~10^{-1}m_{\rm pl}$, which means that 
$\sigma_{\rm max}$ is less than $\sim 0.1 m_{\rm pl}$.
In this case, since $\sigma_{\rm max}$ is smaller than 
the scale $f$ by one order of magnitude, 
the back reaction effect due to 
particle production can be neglected.
When $\phi(0)~\mbox{\raisebox{-1.ex}{$\stackrel
     {\textstyle<}{\textstyle \sim}$}}~10^{-6}m_{\rm pl}$, however,
$\sigma$ exceeds the scale of $f$ and the dynamics of 
inflation are altered. Since the effective potential $(\ref{B18})$
is flat in the region of $\sigma~\mbox{\raisebox{-1.ex}{$\stackrel
     {\textstyle>}{\textstyle\sim}$}}~2m_{\rm pl}$, secondary 
inflation takes place by produced fluctuations.
As compared with the case where the spinodal effect is
neglected, the number of $e$-foldings become
 much larger.
The secondary inflation continues for a long time
in the case of $\phi(0)~\mbox{\raisebox{-1.ex}{$\stackrel
     {\textstyle<}{\textstyle \sim}$}}~10^{-7}m_{\rm pl}$.
Once inflaton exceeds the value of $\phi_0=\pi f/2$,
it gradually approaches the potential minimum
around $\phi_0=\pi f$ and $\sigma \approx 0$,
after which the inflationary period terminates.

If we change two mass scales of $m$ and $f$,
we obtain the smaller maximum fluctuation 
with the decrease of $m$ and $f$ for the fixed 
initial values of $\phi(0)$.
However, if we choose smaller values of $\phi(0)$
which are close to zero, we find that  
fluctuations can grow beyond the perturbative level 
to lead to the secondary inflation.  The number of
$e$-folding depends on the scale of $m$ for the fixed values of
$f$ and $\phi(0)$ if the spinodal effect is taken into account.

We should comment on some points.
The influence due to resonant particle production
on cosmic background anisotropies was studied by several authors
in the context of preheating\cite{pre} and fermion production
during inflation\cite{CKRT} in the chaotic inflation model.
In the new inflation scenario, it was found that  
metric perturbations of super-horizon modes are 
enhanced by spinodal instability 
when inflaton is initially located around $\phi=0$\cite{Boy2}.
It is interesting to study how scale invariance of 
the Harrison-Zel'dovich spectrum would be modified in the 
present model by making use of the gauge invariant 
formalisms of metric perturbations\cite{MP,SMP}.

As related to the treatment of the back reaction by produced particles,
we relied on the Hartree approximation, which is essentially
the mean field approximation.
In the present model, this has the advantage of being able to 
deal with higher order contributions of fluctuations analytically
beyond perturbation theory.
On the other hand, there are other approaches related to back reaction
issues. One of them is 2PI formalism by
Cornwall et al\cite{2PI}. Another approach is to add the stochastic 
noise term due to quantum fluctuations to the field equation,
which is based on the closed time path formalism\cite{stoc}.
These formalisms may alter quantitative details obtained in this
paper especially when fluctuations are enhanced significantly.

Although we studied the natural inflation model as
one example of spinodal inflation, the nonperturbative evolution
of fluctuations which leads to secondary inflation 
would be expected to occur in other spinodal models.
As well as in the ``small field" models such as the natural 
inflation and new inflation, in which 
inflaton is initially small, potentials with spinodal instability
appear in the ``large field" models such as  the higher curvature
gravity\cite{HC} and the nonminimally coupled scalar field
\cite{NM} by performing conformal transformations to 
the Einstein frame. It is of interest how the dynamics of inflation 
are modified in these models by taking into account the effect of
spinodal instability. These issues are in consideration.

\section*{ACKNOWLEDGMENTS}
The authors would like to thank Kei-ichi Maeda for useful discussions.
T. T. is thankful for financial support from the JSPS. This
work was supported partially by a Grant-in-Aid for  Scientific
Research Fund of the Ministry of Education, Science and Culture
(No. 09410217), by a
JSPS Grant-in-Aid (No. 094162), and by the Waseda University Grant 
for Special Research Projects.

\vspace{0.5cm}
{\it Note added.}

Very recently, Cormier and Holman\cite{SP2} considered
the dynamics of the spinodal instability in the same model as ours.
Their results are consistent with our results obtained in this paper.

\newpage

\newpage

\begin{table}
\caption{The number of $e$-folding $N$ and the maximum
value of the fluctuation $\sigma_{\rm max}$ for 
various initial values of $\phi_0$ in the case of
$f=10^{19}$ GeV$ \sim m_{\rm pl}$
and $m=10^{16}$ GeV$ \sim 10^{-3}m_{\rm pl}$.
We also attach the number of $e$-folding $N'$
where the spinodal effect is neglected.
$N$ and $\sigma_{\rm max}$
both increase with the decrease of $\phi(0)$
in the case of $10^{-6}<\phi(0)/m_{\rm pl}
<10^{-1}$.
When $\phi(0)/m_{\rm pl}<10^{-6}$, the fluctuation
grows significantly as $\sigma_{\rm max}/m_{\rm pl}>1$.
This leads to secondary inflation, and the amount of inflation
drastically increases in the case of 
$\phi(0)/m_{\rm pl}<3 \times 10^{-7}$.
}

\vskip .3cm
\noindent
\begin{tabular}{clcccccc}
        ~& $\phi(0)/m_{\rm pl}$ & ~& $N$ &~
        & $N'$ 
        & $\sigma_{\rm max}/m_{\rm pl}$ 
        &~\\
        \hline
~&$1 $ &~& 39 &~& 39 & $-$ &~\\
~&$ 0.5 $ &~& 71 &~& 71 & $-$ &~\\       
~&$1 \times 10^{-1}$ &~& 156 &~& 156 & $1.1 \times 10^{-5}$ &~\\
~&$1 \times 10^{-2}$ &~& 272 &~& 272 & $1.1 \times 10^{-4}$ &~\\
~&$1 \times 10^{-3}$ &~& 387 &~& 387 & $1.1 \times 10^{-3}$ &~\\
~&$1 \times 10^{-4}$ &~& 502 &~& 502 & $1.1 \times 10^{-2}$ &~\\
~&$1 \times 10^{-5}$ &~& 618 &~& 618& $1.1 \times 10^{-1}$ &~\\
~&$1 \times 10^{-6}$ &~& 776 &~& 734 & 1.1&~\\
~&$5 \times 10^{-7}$ &~& 1151 &~& 771 & 2.3 &~\\
~&$3 \times 10^{-7}$ &~& 22178 &~& 797 & 3.8 &~\\
\end{tabular}
\end{table}

\begin{table}
\caption{The number of $e$-folding $N$ and the maximum
value of the fluctuation $\sigma_{\rm max}$ for 
various initial values of $\phi_0$ in the case of
$f=10^{19}$ GeV$ \sim m_{\rm pl}$
and $m=10^{15}$ GeV$ \sim 10^{-4}m_{\rm pl}$.
We also attach the number of $e$-folding $N'$
where the spinodal effect is neglected.
Compared with the case of 
$f=m_{\rm pl}$ and $m=10^{-3}m_{\rm pl}$,
$\sigma_{\rm max}$ is smaller by two orders of magnitude
for the same initial values of $\phi$
while $N$ is the same value for $\phi(0)/m_{\rm pl}
>10^{-5}$. In this case, secondary inflation takes place for 
$\phi(0)/m_{\rm pl}<5 \times 10^{-9}$, which is 
smaller than in the case of 
$f=10^{19}$ GeV and $m=10^{16}$ GeV.
}

\vskip .3cm
\noindent
\begin{tabular}{clcccccc}
        ~& $\phi(0)/m_{\rm pl}$ & ~& $N$ &~
        & $N'$ 
        & $\sigma_{\rm max}/m_{\rm pl}$ 
        &~\\
        \hline
~&$1 \times 10^{-1}$ &~& 156 &~& 156 & $-$ &~\\
~&$1 \times 10^{-2}$ &~& 272 &~& 272 & $-$ &~\\
~&$1 \times 10^{-3}$ &~& 387 &~& 387 & $1.4 \times 10^{-5}$ &~\\
~&$1 \times 10^{-4}$ &~& 502 &~& 502 & $1.4 \times 10^{-4}$ &~\\
~&$1 \times 10^{-5}$ &~& 618 &~& 618 & $1.4 \times 10^{-3}$ &~\\
~&$1 \times 10^{-6}$ &~& 734 &~& 734& $1.4 \times 10^{-2}$ &~\\
~&$1 \times 10^{-7}$ &~& 851 &~& 851 & $1.4 \times 10^{-1}$ &~\\
~&$1 \times 10^{-8}$ &~& 1043 &~& 967 & 1.4&~\\
~&$7 \times 10^{-9}$ &~& 1234 &~& 986 & 2.0 &~\\
~&$5 \times 10^{-9}$ &~& 2512 &~& 1004 & 2.9 &~\\
\end{tabular}
\end{table}

\begin{table}
\caption{The number of $e$-folding $N$ and the maximum
value of the fluctuation $\sigma_{\rm max}$ for 
various initial values of $\phi_0$ in the case of
$f=5.0 \times 10^{18}$ GeV$\sim 0.5m_{\rm pl}$ and
$m=10^{15}$ GeV$\sim 10^{-4}m_{\rm pl}$. 
We also attach the number of $e$-folding $N'$
where the spinodal effect is neglected.
Both $N$ and $\sigma_{\rm max}$ take smaller values
compared with the case of 
$f=10^{19}$ GeV and $m=10^{15}$ GeV
for the same initial value of $\phi$.
Secondary inflation occurs for initial values of 
$\phi(0)/m_{\rm pl}<5 \times 10^{-9}$.
}

\vskip .3cm
\noindent
\begin{tabular}{clcccccc}
        ~& $\phi(0)/m_{\rm pl}$ & ~& $N$ &~
        & $N'$ 
        & $\sigma_{\rm max}/m_{\rm pl}$ 
        &~\\
        \hline
~&$1 \times 10^{-1}$ &~& 33 &~& 33 & $-$ &~\\
~&$1 \times 10^{-2}$ &~& 62 &~& 62 & $-$ &~\\
~&$1 \times 10^{-3}$ &~& 92 &~& 92 & $6.7 \times 10^{-6}$ &~\\
~&$1 \times 10^{-4}$ &~& 122 &~& 122 & $6.7 \times 10^{-5}$ &~\\
~&$1 \times 10^{-5}$ &~& 151 &~& 151 & $6.7 \times 10^{-4}$ &~\\
~&$1 \times 10^{-6}$ &~& 180 &~& 180 & $6.7 \times 10^{-3}$ &~\\
~&$1 \times 10^{-7}$ &~& 210 &~& 210 & $6.7 \times 10^{-2}$ &~\\
~&$1 \times 10^{-8}$ &~& 256 &~& 240 & $6.7 \times 10^{-1}$ &~\\
~&$7 \times 10^{-9}$ &~& 290 &~& 244 & $9.4 \times 10^{-1}$ &~\\
~&$5 \times 10^{-9}$ &~& 458 &~& 249 & 1.3 &~\\
\end{tabular}
\end{table}

\newpage
\begin{flushleft}
{ Figure Captions}
\end{flushleft}
\noindent
\parbox[t]{2cm}{FIG. 1:\\~}\ \
\parbox[t]{8cm}
{The effective two-field potential
$V(\phi_0, \sigma) \equiv
m^4 \left[ 1+ \exp \left(-\frac{\sigma^2}{2f^2}
\right) \cos  \frac{\phi_0}{f} \right]$ in 
the natural inflation model.
$\phi_0$ and $\sigma$ are normalized by the scale $f$.
In the case where  
growth of the fluctuation of inflaton is neglected, this 
is reduced to the one-field potential 
$V(\phi_0)=m^4 \left[1+\cos\left(\frac{\phi_0}{f}\right)
\right]$. However, when the fluctuation grows
significantly and evolves toward the $\sigma$ direction,
the system is described by two fields 
$\phi_0$ and $\sigma$.
}\\[1em]
\noindent
\parbox[t]{2cm}{FIG. 2:\\~}\ \
\parbox[t]{8cm}
{The evolution of $\phi_0$ and $\sigma$ fields
for the initial value of $\phi(0)=0.1m_{\rm pl}$
in the case of $f=10^{19}$ GeV$ \sim m_{\rm pl}$
and $m=10^{16}$ GeV$ \sim 10^{-3}m_{\rm pl}$.
Both fields are normalized by $m_{\rm pl}$.
Although fluctuations grow at the initial stage, 
the maximum value $\sigma_{\rm max}
\approx 10^{-5}m_{\rm pl}$ 
achieved in this case is much smaller than the value $f$.  
The $\phi_0$ field evolves toward the potential 
minimum $\phi_0 = \pi f$,
after which the universe enters the reheating stage.
In this case, the enhancement of fluctuations hardly 
affects the evolution of the $\phi_0$ field and the 
scale factor.
 }\\[1em]
\noindent
\parbox[t]{2cm}{FIG. 3:\\~}\ \
\parbox[t]{8cm}
{The evolution of $\phi_0$ and $\sigma$ fields
for the initial value of 
$\phi(0)=5.0 \times 10^{-7}m_{\rm pl}$ 
in the case of $f=10^{19}$ GeV$ \sim m_{\rm pl}$
and $m=10^{16}$ GeV$ \sim 10^{-3}m_{\rm pl}$.
The fluctuation $\sigma$ reaches the maximum value
$\sigma_{\rm max}=2.3m_{\rm pl}$ when $\phi_0=\pi f/2$.
After that, $\sigma$ decreases because
spinodal instability is absent.
The inflaton field finally rolls down toward the 
potential minimum with $\phi_0=\pi f$ and $\sigma
\approx 0$.
 }\\[1em]
\noindent
\parbox[t]{2cm}{FIG. 4:\\~}\ \
\parbox[t]{8cm}
{The evolution of $\phi_0$ and $\sigma$ fields
for the initial value of 
$\phi(0)=3.0 \times 10^{-7}m_{\rm pl}$ 
in the case of $f=10^{19}$ GeV$ \sim m_{\rm pl}$
and $m=10^{16}$ GeV$ \sim 10^{-3}m_{\rm pl}$.
The fluctuation $\sigma$ reaches the maximum value
$\sigma_{\rm max}=3.8m_{\rm pl}$ 
when $\phi_0=\pi f/2$.
The secondary inflation due to fluctuations
occurs for $1 \times 10^{6}~\mbox{\raisebox{-1.ex}{$\stackrel
     {\textstyle<}{\textstyle \sim}$}}~mt~\mbox{\raisebox{-1.ex}{$\stackrel
     {\textstyle<}{\textstyle \sim}$}}~7 \times 10^{6}$.
In this region, since the effective two-field potential
is very flat, the inflaton field evolves very slowly.
Finally, inflaton is trapped in the potential minimum
$\phi_0=\pi f$ and $\sigma \approx 0$ at
$mt=7.6 \times 10^6$.
 }\\[1em]
\noindent
\parbox[t]{2cm}{FIG. 5:\\~}\ \
\parbox[t]{8cm}
{The evolution of the Hubble parameter $H$
for the initial value of 
$\phi(0)=3.0 \times 10^{-7}m_{\rm pl}$ 
in the case of $f=10^{19}$ GeV$ \sim m_{\rm pl}$
and $m=10^{16}$ GeV$ \sim 10^{-3}m_{\rm pl}$.
The first inflation occurs around $\phi_0 \approx 0$
for $0~\mbox{\raisebox{-1.ex}{$\stackrel
     {\textstyle<}{\textstyle \sim}$}}~mt~\mbox{\raisebox{-1.ex}{$\stackrel
     {\textstyle<}{\textstyle \sim}$}}~1.3 \times 10^{5}$, which is 
followed by the secondary inflation caused by fluctuations
for $1 \times 10^{6}~\mbox{\raisebox{-1.ex}{$\stackrel
     {\textstyle<}{\textstyle \sim}$}}~mt~
     \mbox{\raisebox{-1.ex}{$\stackrel
     {\textstyle<}{\textstyle \sim}$}}~7 \times 10^{6}$.
Since the duration of this secondary inflation is long,
the achieved number of $e$-folding is very large
as $N=22178$.
 }\\[1em]
\noindent

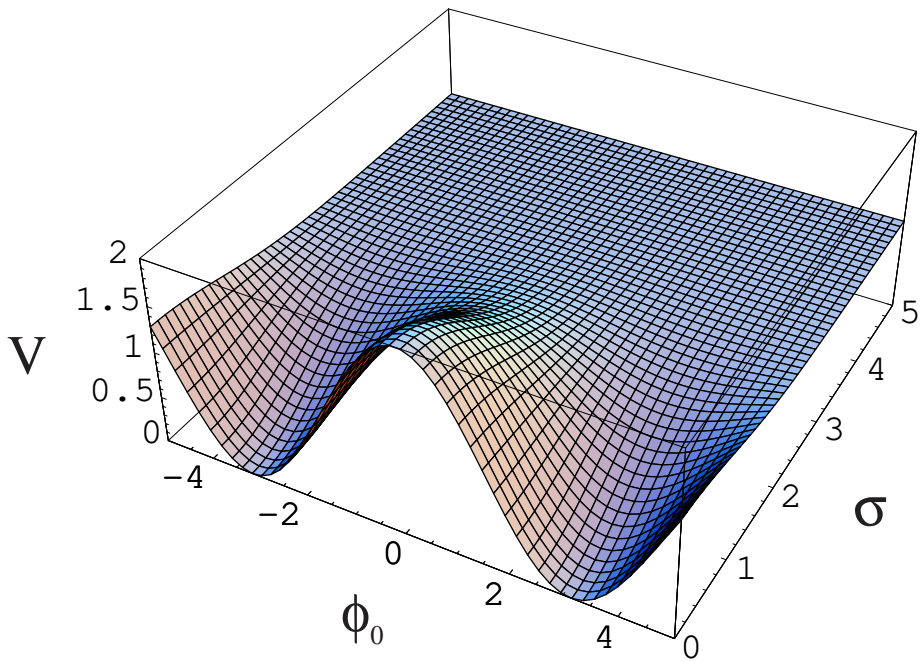
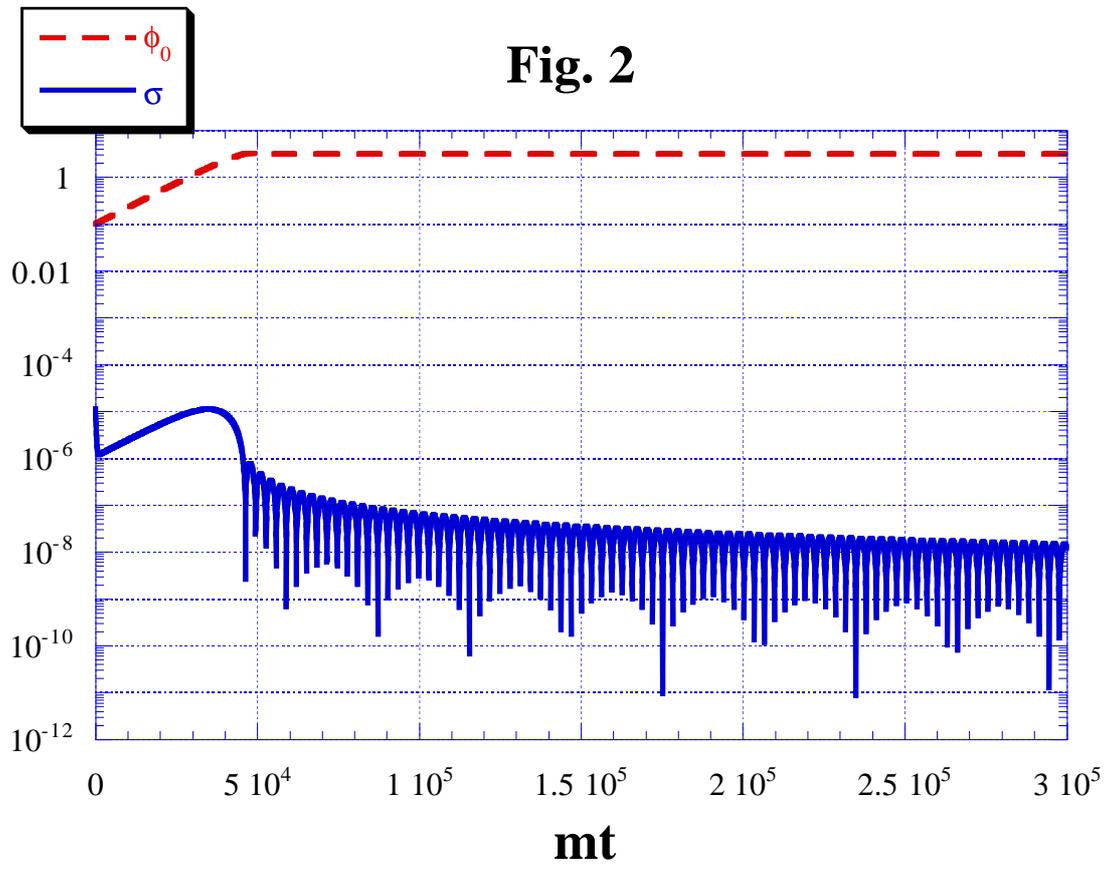
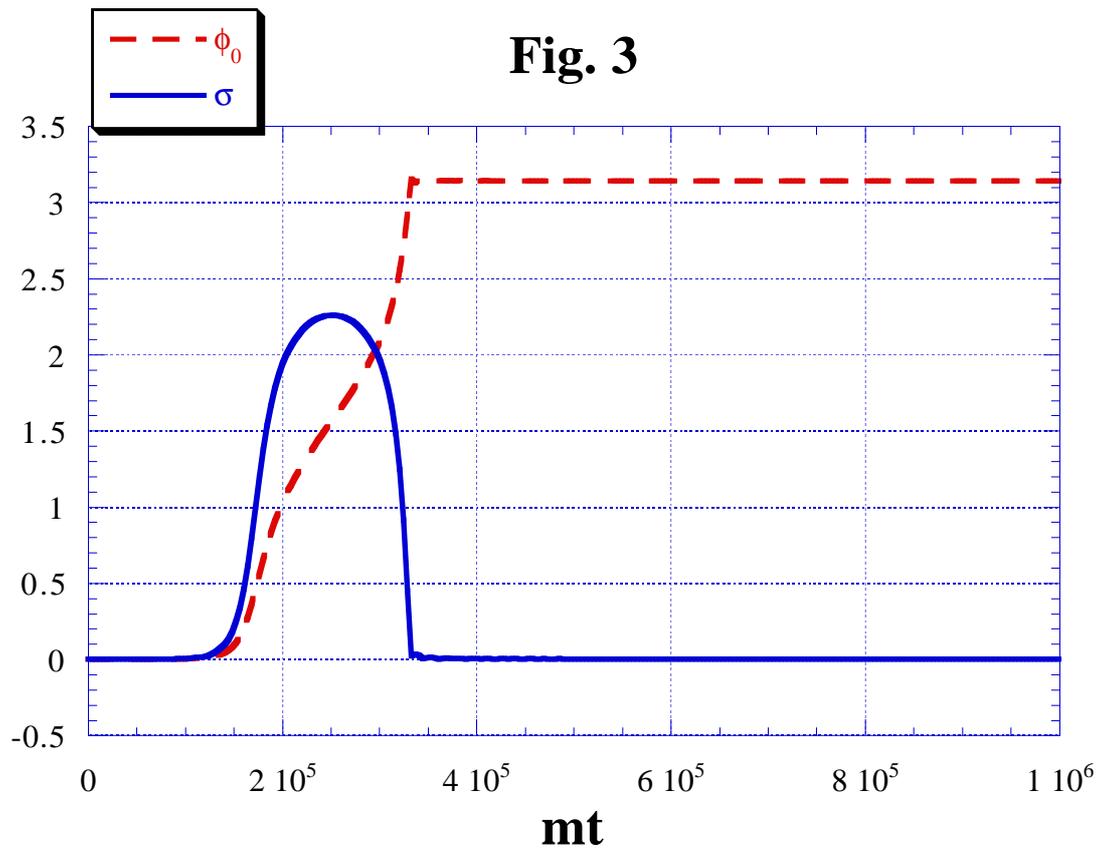
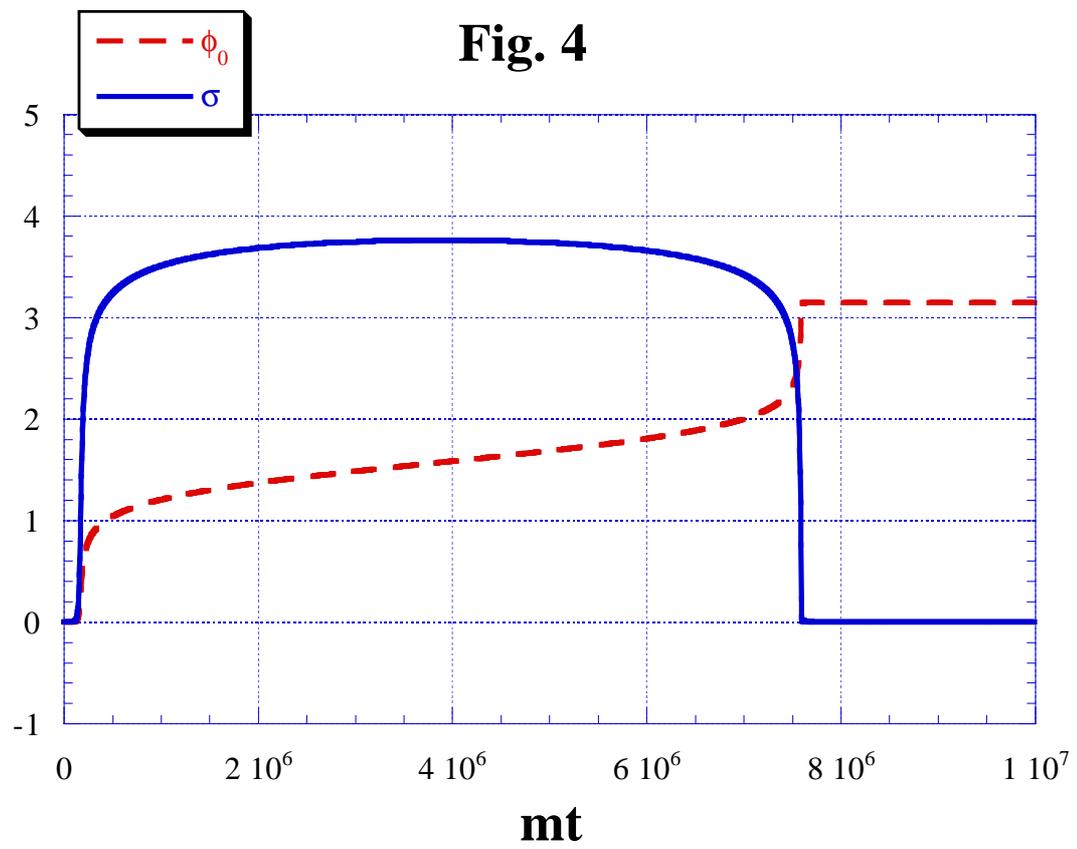
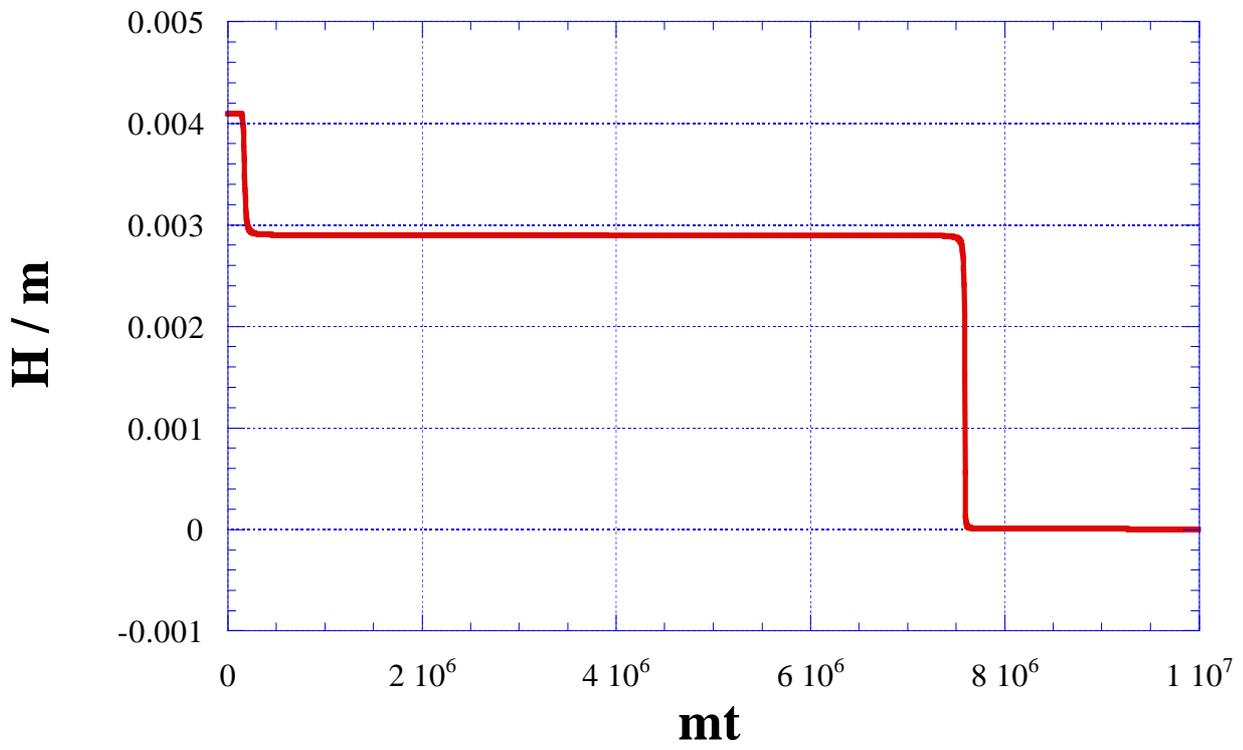
\end{document}